# Magnetic hot-spots generation at optical frequencies in all-dielectric mesoscale Janus particles

Oleg V. Minin[a], Song Zhou[b], and Igor V. Minin[a]
[a]Tomsk Polytechnic University, 30 Lenin Ave., Tomsk 634050 Russia
[b]Jiangsu Key Laboratory of Advanced Manufacturing Technology, Faculty of Mechanical and Material Engineering, Huaiyin Institute of Technology, Huai'an 223003, China.

***Abstract***. At optical frequencies due to the small value of the magnetic permeability of natural materials, the magnetic effects are week. To this end, the natural dielectric materials are unemployable for practical "magnetic" applications in optics. We have shown that it is possible to induce the intense magnetic hot spots in a Janus dielectric mesoscale particle. The basic idea of the Janus particle based on a combination of the effects of a photonic jet, whispering gallery waves and the concept of solid immersion. Simulations show that $H^2/E^2$ contrast maybe more 10 and maximal magnetic field intensity enhancement is more than 1000 for a wavelength-scaled particle with refractive index n < 2.

Strong localization of optical wave to volumes below the diffraction limit is a topic of extensive research involving a wide range of applications [1–6]. The ability to localize optical wave in sub-wavelength volumes are called hot-spots [7], at which the electric and even magnetic fields intensities are enhanced up to several orders of magnitude.

A single dielectric spherical high-permittivity nanoparticle, can exhibit both electric and strong magnetic resonances [8] where the first fundamental mode corresponds to a magnetic dipole excitation, but fabrication tolerances must be tailored down to the sub-nm resolution. The generation of strong magnetic hot-spots by dielectric nanoparticles was observed in the inter-particle regions [9]. The interference of magnetic and electric modes in such nanoparticles assemblies give rise to sharp magnetic Fano resonances [10-13].

Dielectric wavelength-scaled (mesoscale) particles with Mie size parameter *q=kR (k-* wavenumber and *R* – particle radius*)* in order of *q ~10* have aroused big interest because of their potential to localize light at subwavelength scale [14] and to yield high internal magnetic and electric local field enhancements instead of plasmonic metal nanoparticles [15-17]. On the other hand, the employment of mesoscale dielectric particles has facilitated the achievement of a remarkable magnetic enhancement overcoming the inherent losses of plasmonic materials. The optical magnetic fields localization squeezinging in deep sub-wavelength regions, opening promising perspectives for spintronics.

In [18] it has been shown that Mie type resonances of different orders overlap with increasing the refractive index much more than 1 which lead to a high concentration of the electric and magnetic fields within the dielectric spherical low loss particle with diameter less than wavelength. For spherical gallium phosphide particle with refractive index n=3.4932 at the wavelength of 532 nm and with Mie size parameter of *q*=5.38 the maximal fields intensity of $E^2$~40 and $H^2$~140 were observed.

In [19] Haong et al. reported high-resonance effect using a particle with *q*~37 and *n*=1.46. It was observed that one of the resonant scattering coefficient was 20 time higher in magnitude than the other coefficients. This abnormal value of the scattering coefficient was described as the constructive interference of the one partial wave inside the microsphere. Later, the optical so-called "super-resonance effect" in mesoscale dielectric spheres based on the high-order Fano resonance and caused by particle's internal partial waves was offered in [20-22] which theoretically allow to achieving super-high intensity of magnetic fields.

This effect is valid for the certain size parameters $q \sim 10...70$ and the refractive index $n < 2$, which theoretically rendered the field enhancement more than $10^7$ times stronger than that of the downward radiation [23]. On the other hand, it demonstrated the possibility of overcoming the diffraction limit despite the high sensitivity to losses in the particle material. An unusual effect - the hotspots size decrease down to less than the diffraction limit (or even less than in the ideal case of the particle material without losses) after the introduction of small dissipation into the particle material - was observed for the first time in [22].

While spherical mesoscale dielectric particle shapes have only 2 degrees of freedom (Mie size parameter $q$ and refractive index of the particle material), optically asymmetric particles (particles with broken spherical or cylindrical symmetries, so-called Janus particle [24]) provide additional degrees of flexibility in electromagnetic response tuning [25]. While shaping the high-order Fano resonance has opened opportunities for localized magnetic and electric fields manipulation, in [25] we have propose a more fundamental approach. By tailoring the broken symmetry of the spherical or cylindrical particles shape, we can facilitate new kinds of localization and enhancement of the electromagnetic fields hot spots inside a Janus particle near its shadow surface. Introduction of broken symmetry into dielectric spherical or cylindrical particles as additional degree of freedom enlarges the capabilities of strong field's localization beyond diffraction limit at the nanoscale, opening a room of opportunities for new possible applications. In this manner, we find that spherical or cylindrical dielectric mesoscale particles with broken symmetry can generate stable nanoscale hot spots with giant field intensity enhancement [25].

The main idea of new mechanism of the field localization in the Janus particle proposed in [25] is a combination of the effects of a photonic nanojet and whispering gallery waves. At a fixed truncation height $h$, sharp resonances are observed in the intensities of the electric and magnetic fields as a function of the Mie size parameter $q$. Approximately the same distributions of resonances is observed in the case of high Fano resonances [20]. With a change in the truncation thickness $h$, a narrow resonance is observed for the TM mode, when the vector H of the incident plane wave lies in the *x-y* plane of the truncated element of the sphere. In this case, hot spots with extremely high intensity appear on the flat surface of the truncated element, which associated with the excitation of whispering gallery waves on a flat element of a truncated surface [25].

Let's consider the cylindrical particle with the following main parameters: the radius of the cylinder R= 5λ (correspond to the resonant size parameter of q=31.41593) at wavelength of λ=500 nm and refractive index of the particle material as $n_p$=1.5. These are the usual particle parameters for the formation of a photonic jet. Below we will use a refractive index contrast ($n=n_p/n_c$) between the particle $n_p$ material in cutting area $n_c$.

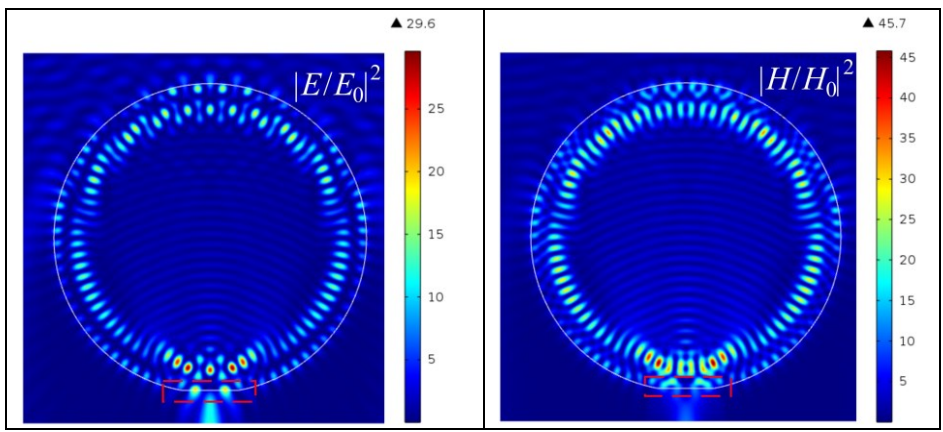

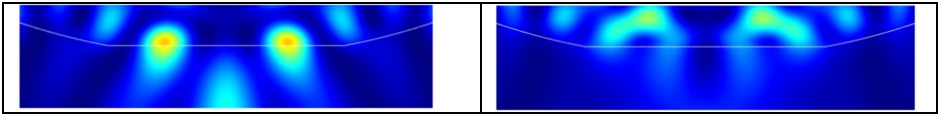

Fig.1. Hot spots generation in Janus cylindrical particle with parameters of h=42 nm, n=1.5 (refractive index of the cutting area $n_c$=1, vacuum).

Figure 1 clearly shows the "electric" photonic jet and two hot spots near the flat boundary of the Janus particle. Detailed studies of the resonance properties of such a Janus particle are given in [25].

The work of the Janus particle can be clearly explained based on the geometric-optical approximation [26]. When radiation is incident at an angle of total internal reflection $\chi_0 = \arcsin(n_m/n_p)$, the flat surface plays the role of a mirror. The interference of waves incident on a flat surface at angles of total internal reflection creates high-intensity evanescent fields near the surface. For example, at n = 1.5, $\chi_0 \cong 41.8^o < 45^0$, but for small truncations of a cylindrical particle, the first resonance should be observed at $\chi \approx \frac{\pi}{4} = 45^o$. The difference in the angle values $\chi$ is explained by the fact that on the line of intersection of a flat section with a cylindrical surface, a wave phase jump occurs, which can be determined from the generalized law of refraction (see Fig. 2) [27, 26]:

$$n_p \sin\theta_p - n_m \sin\theta_m = \frac{\lambda}{2\pi}\frac{d\Phi}{dx}. \qquad (1)$$

where $d\Phi/dx$ is the change in the phase gradient of the wave, depending on the thickness of the truncated element h. In this case, the angle of total internal reflection changes as [26]:

$$\chi = \arcsin(\frac{n_m}{n_p} + \frac{\lambda}{2\pi n_p}\frac{d\Phi}{dx}). \qquad (2)$$

Note that for small truncations h, the correction $\Delta\chi$ to the angle of total internal reflection $\chi_0$ is proportional to the thickness of the truncated element h and inversely proportional to the refractive index $n_p$: $\chi \approx \chi_0 + \Delta\chi$, $\Delta\chi = \beta\frac{h}{n_p q}$, $\beta = const$ (Fig.2).

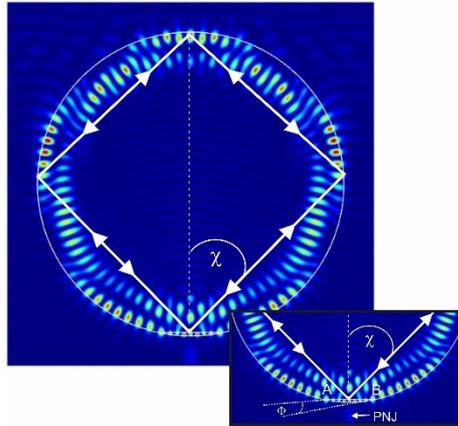

Fig.2. The path of rays in a Janus particle in the case of a ray falling on a flat surface at an angle of total internal reflection $\chi$. The inset shows a schematic change in the phase of a wave along a section of a flat surface.

The development of the Janus particle concept [25] consists in the involvement of the third effect - the concept of solid immersion integrated onto a dielectric particle. The particle cutting into two parts, which has different refractive index material in main lower and small upper parts. The high-index material in small upper part allows new Janus particle accessing contribution from solid immersion mechanism [28].

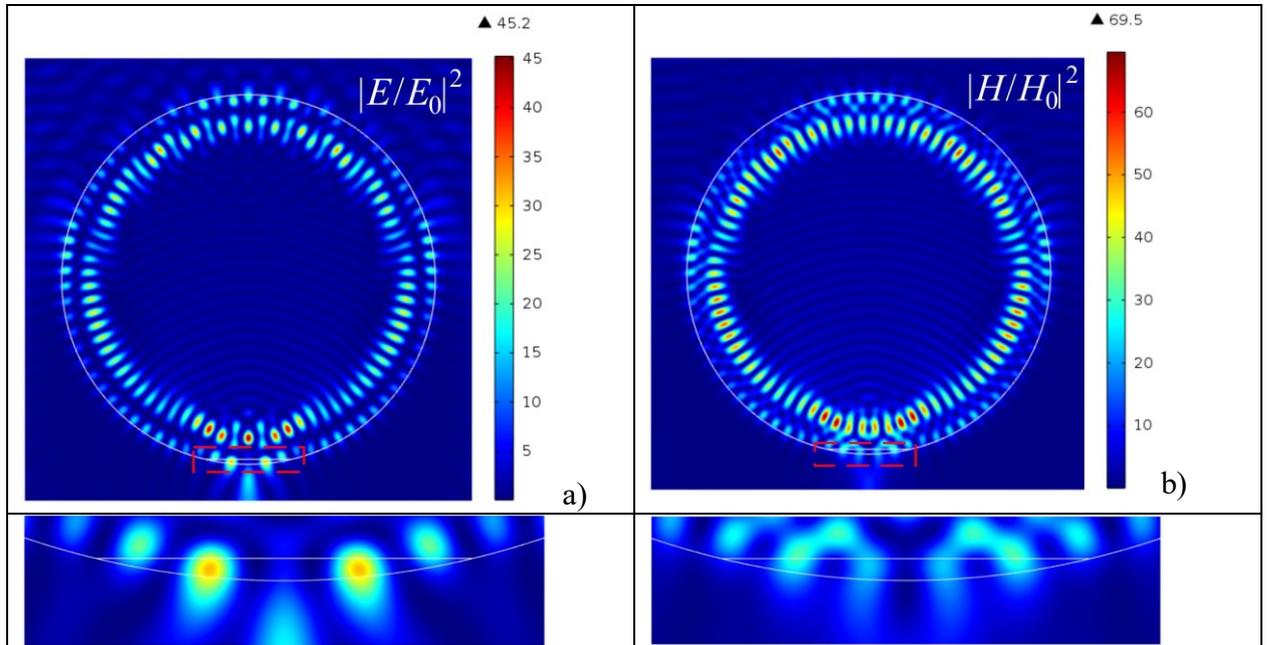

Fig.3. Hot spots generation in Janus cylindrical particle with parameters of h=68 nm, n=1.1236 (water).

Figure 3 shows the generation of hot spots when the truncated portion of the cylinder is filled with water. Since the refractive index of water is intermediate between the refractive index of a particle and vacuum, this part of the particle acts as a dielectric matching layer that reduces reflection from a flat surface [29]. In the Figure 3 it is also clearly seen that an "electric" photonic jet is formed in the shadow part of the Janus particle.

The formation of a photonic jet in this case is due to the specific distribution of hot spots and vortices [23] inside the particle and the low index dielectric layer in its shadow part, which is shown in Figure 3c.

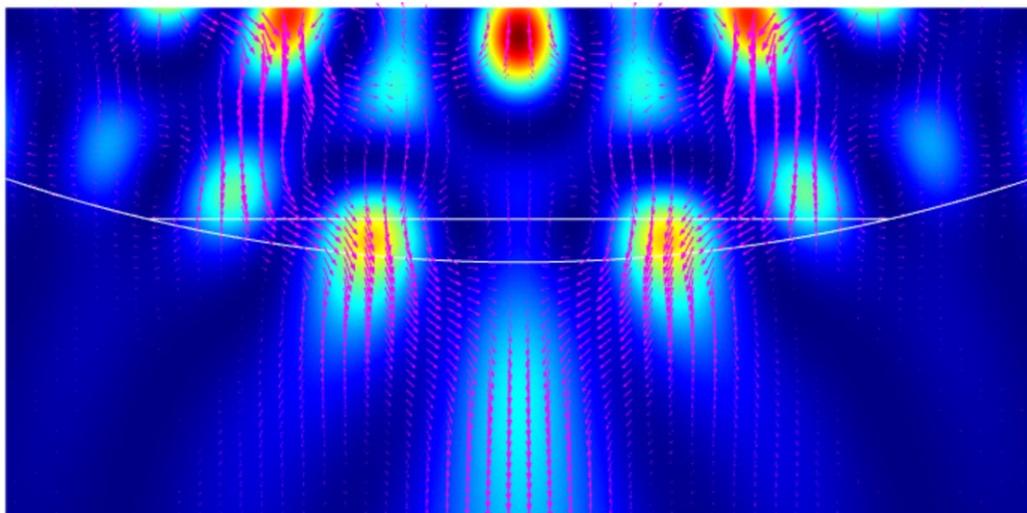

Fig.3c. Distribution of the Poynting vector around the hotspots of the Janus particle.

With an increase in the contrast of the refractive index and, in the dielectric layer with a high refractive index, zones of hot spots with high intensity are formed. This situation is shown in Figure 4.

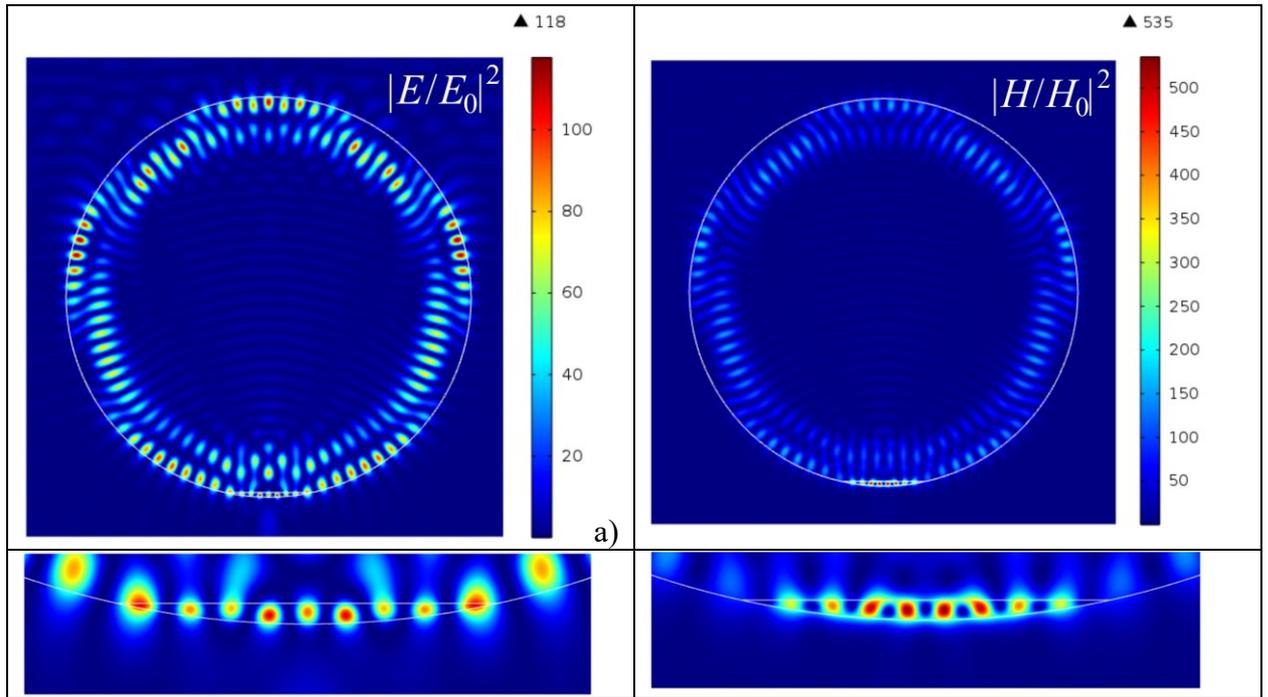

Fig.4. Hot spots generation in Janus cylindrical particle with parameters of h=58 nm, n=0.476.

By comparing field intensity distribution of Figure 1 and Figure 4, two-materials composite Janus cylindrical particle has more converged hot spots than initial (Fig.1) configuration. Moreover, one can see that multiple localized hot spots are in an annular arrangement across cylindrical boundary (due to cylindrical symmetry of the Janus particle) and several highest enhancement appear inside of internal high-index material, causing by wave interferences at two material interfaces.

Figure 5 shows the field intensity distribution along the extrema of the hot spots for electric (a) and magnetic (b) components. Figure 5c shows the vortices and the Poynting vector energy flux near the hot spots, demonstrating complex vortex flow in this area.

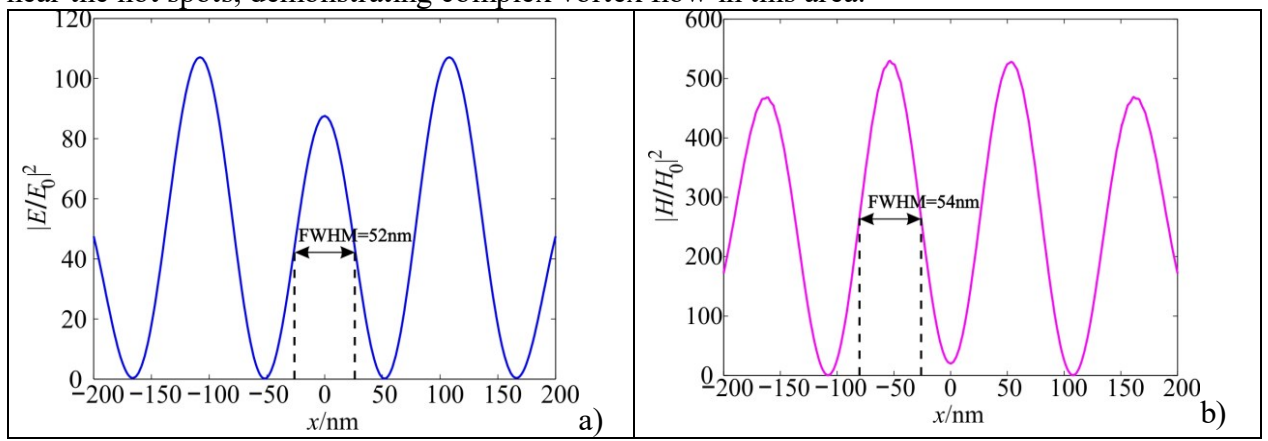

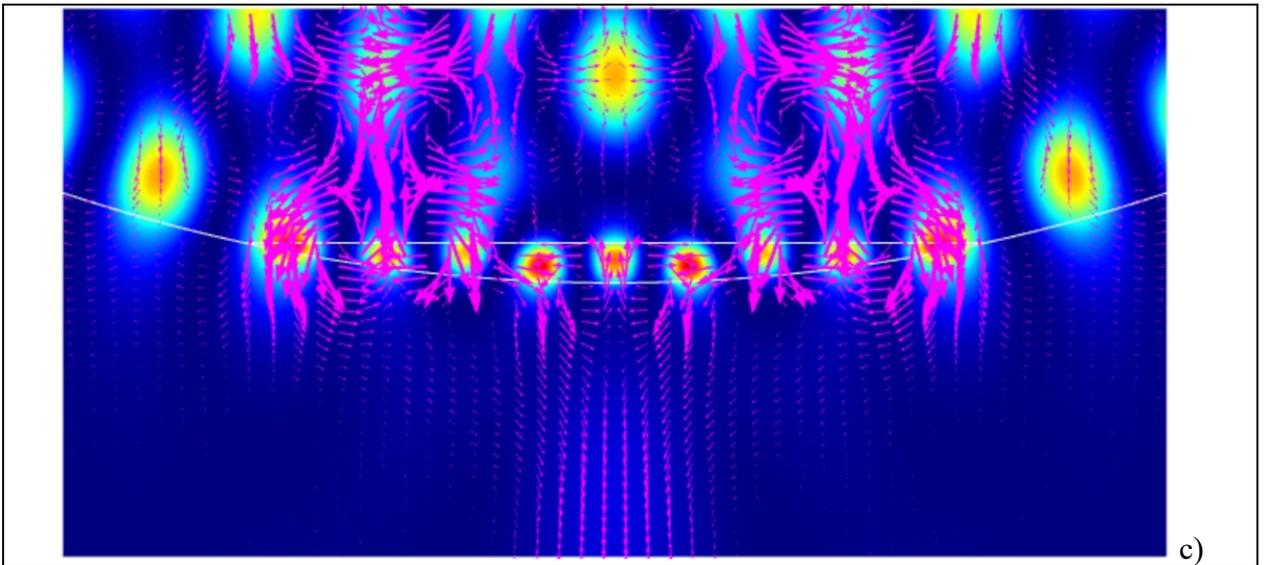

Figure 5. Field intensity distribution along the extrema of the hot spots for electric (a) and magnetic (b) components and the vortices and the Poynting vector energy flux (c).

One can see that the half-width of the intensity maximum for both electric and magnetic fields is about $0.11\lambda$, which is much smaller than the solid immersion limit criterion. In this case, the enhancement of the intensity of the magnetic field is about 500, which is about 4-5 times higher than that for the electric field.

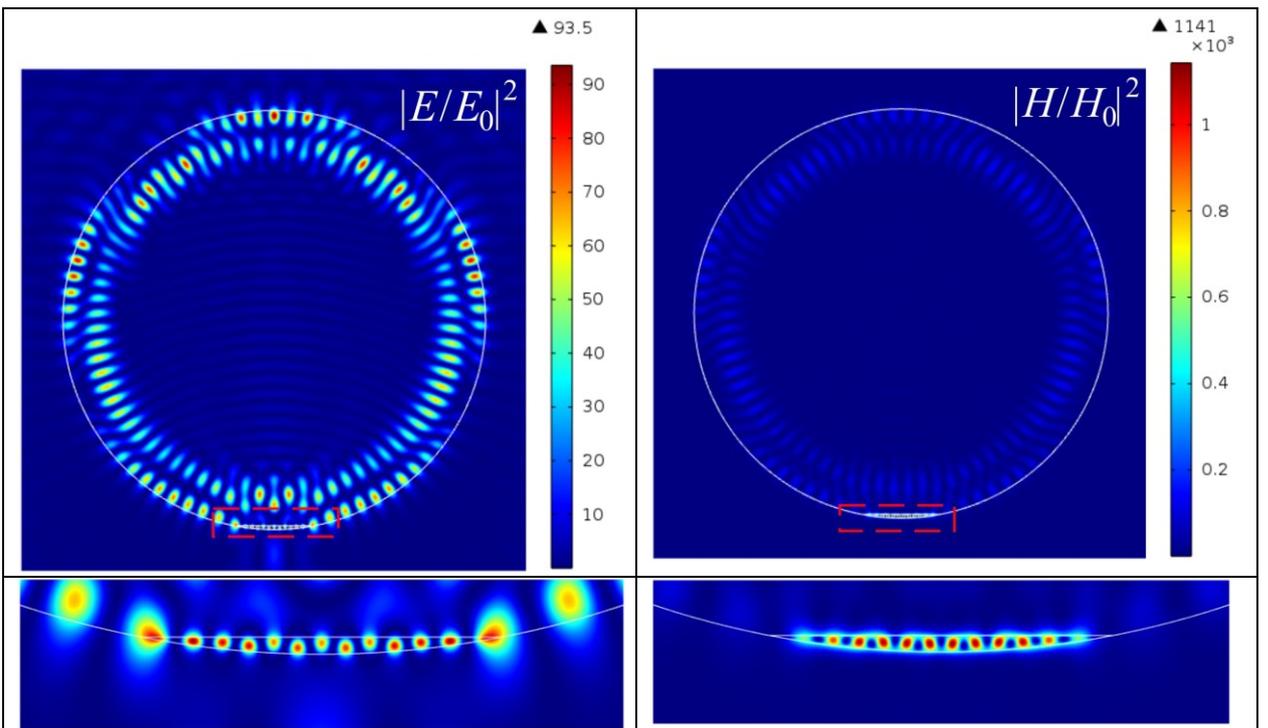

Figure 6. Hot spots generation in Janus cylindrical particle with parameters of h=46 nm, n=0.3.

A further increase in the refractive index of the material of the truncated cylinder leads to an even greater increase in the intensity of the hot spots of the magnetic field, which are shown in Figs.6-7. One can see that the half-width of the intensity maximum for both electric and magnetic fields is about $0.064\lambda$, which is also much smaller than the solid immersion limit criterion. In this case, the enhancement of the intensity of the magnetic field is about 1000, which is about 12 times higher than that for the electric field.

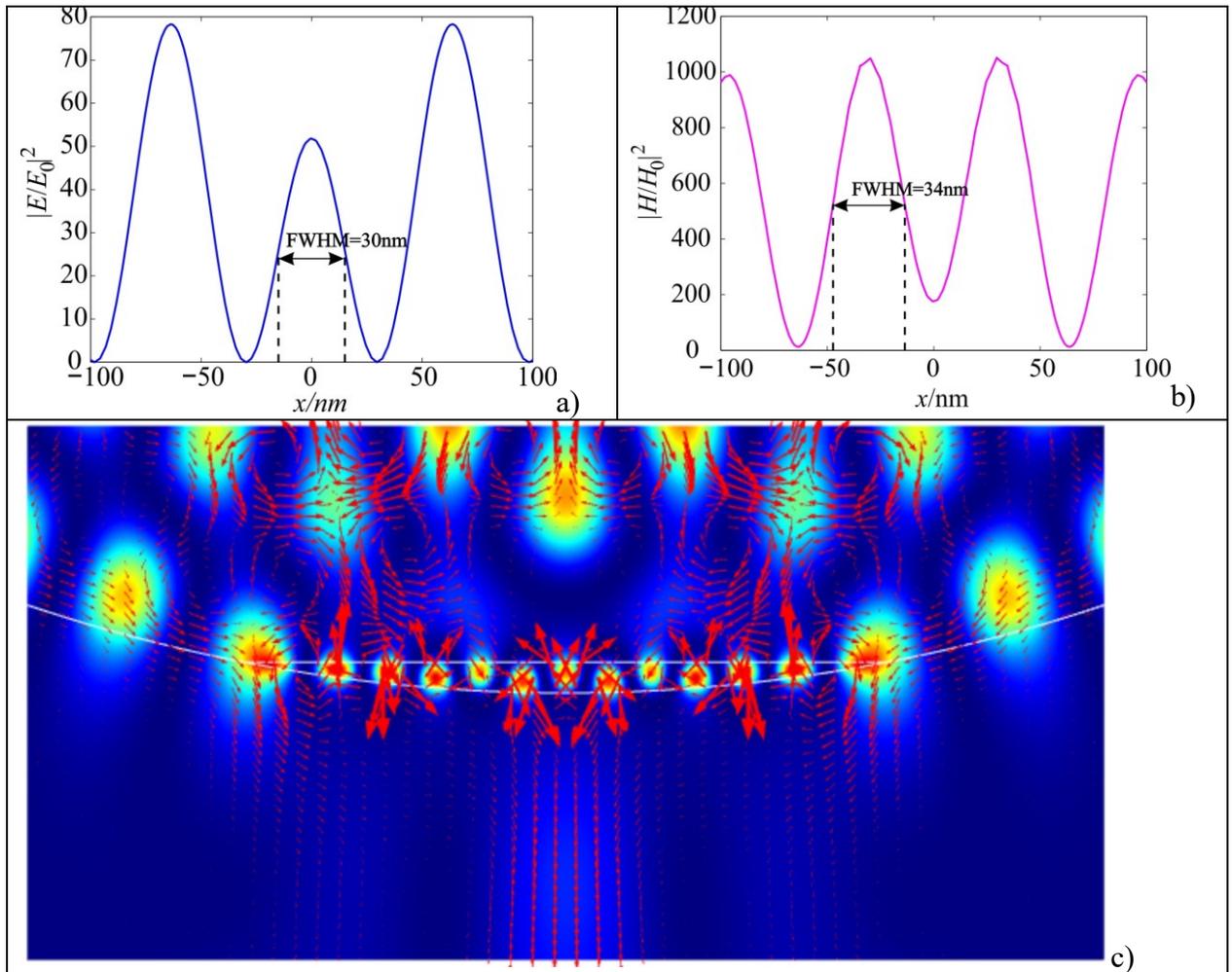

Figure 7. Field intensity distribution along the extrema of the hot spots for electric (a) and magnetic (b) components and the vortices and the Poynting vector energy flux (c).

It is known that so-called super resonances [20-23,30] are quite sensitive to dissipation and with a low dissipation these resonances are strongly suppressed. Below in Fig.8 we show magnetic hot spots generation for particle with ref index contrast of n=0.3 as in Fig.5 but for material of dielectric at bottom side as Rhenium diselenide (ReSe$_2$) having ref index near 5 and losses k=0.005 [31].
Note that the values of the magnetic field intensity for n=0.3 are approximately two times higher than those for spherical particles [18] without losses.

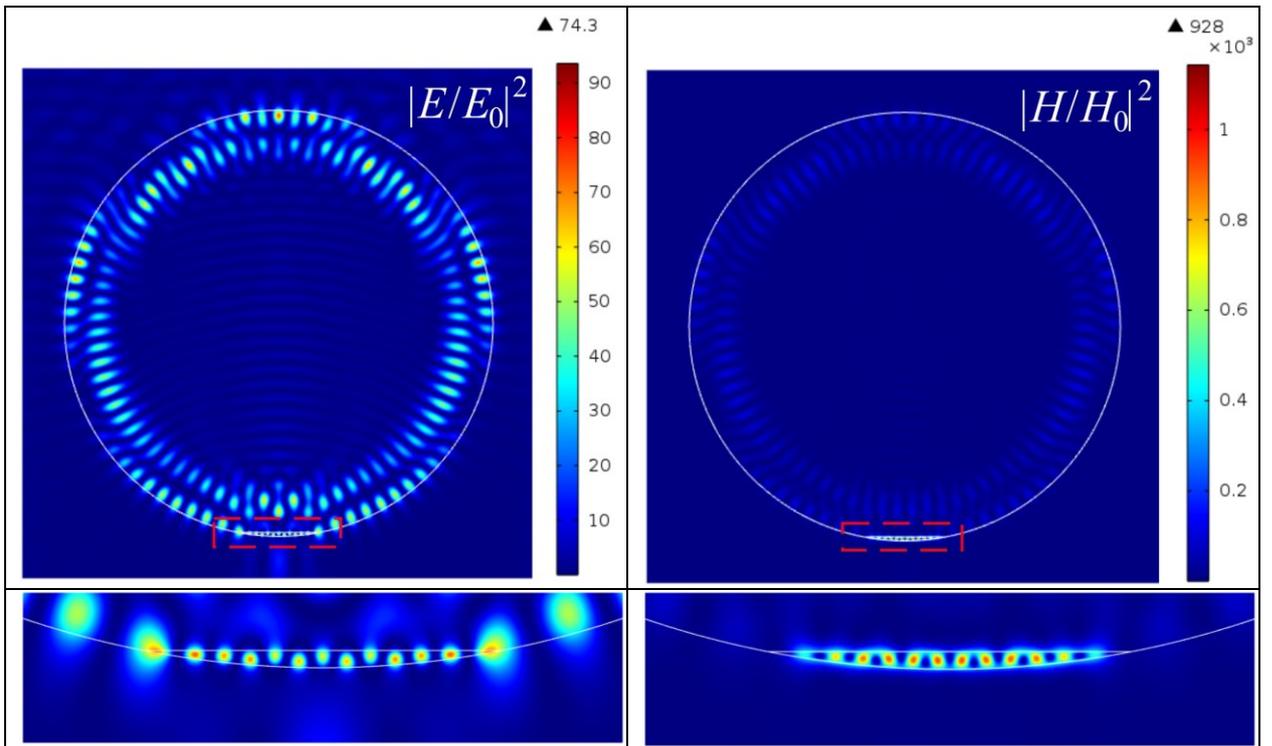

Figure 8. Hot spots generation in Janus cylindrical particle with parameters of h=46 nm, n=0.3 and k=0.005.

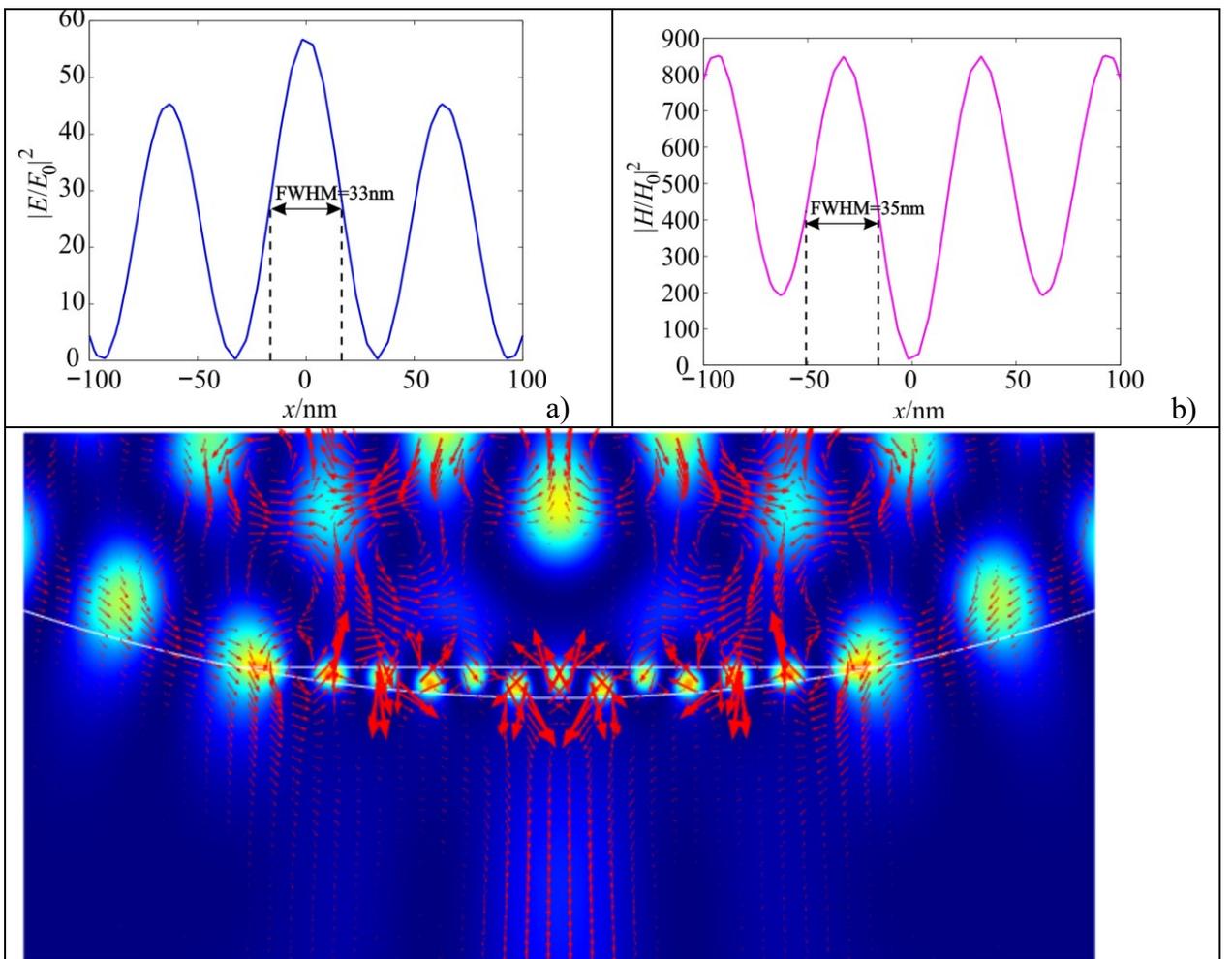

Figure 9. Field intensity distribution along the extrema of the hot spots for electric (a) and magnetic (b) components and the vortices and the Poynting vector energy flux (c).

Comparative characteristics of hot spots of Janus particles are presented in the Table.

Table.

| n | $(H/H_0)^2$ | $(E/E_0)^2$ | $(H/H_0)^2/(E/E_0)^2$ |
|---|---|---|---|
| 1.5 | 45 | 29 | 1.55 |
| 1.124 | 69 | 45 | 1.53 |
| 0.476 | 535 | 118 | 4.53 |
| 0.3 (k=0) | 1141 | 93 | 12.27 |
| 0.3 (k=0.005) | 928 | 74 | 12.54 |

The introduction of losses into the dielectric material led to a drop in the intensity of the electric field by almost 20%, magnetic - by 18%.

**Conclusion**

Generation of deep subwavelength magnetic hot spots, based on a new physical principles, aside from their key role in fundamental physics, provide a new degree of freedom for all-dielectric mesoscale structures, which can control unconventional photonic processes. Thus, artificial optical magnetism is an active topic of research, and great attention has been devoted to all dielectric wavelength-scaled structures generating magnetic hot spots. We have shown that it is possible to induce the intense magnetic hot spots in a Janus dielectric mesoscale particle. The basic idea of the Janus particle based on a combination of the effects of a photonic jet, whispering gallery waves and the concept of solid immersion. Applying a morphological symmetry breaking on the cylindrical particle, we could switch from electric field hot spots to a magnetic one with field enhancement up to several orders of magnitude. As expected, magnetic and electrical hot spots are sensitive to losses in the dielectric material.
Simulations show that $H^2/E^2$ contrast maybe more 10 for a wavelength-scaled particle with refractive index n < 2 and it increases with the increase of Mie size parameter. For such Janus particles a conventional nonlinear optics related to nonlinearity $\varepsilon = \varepsilon(E)$ is dominant.

**Acknowledgements**